\documentclass[aps,showpacs,prb,twocolumn,supperscriptaddress]{revtex4-1}
\usepackage{amsbsy,amssymb,amsmath,bm}
\usepackage{graphicx,color,epsfig,rotate}
\usepackage{epstopdf}
\usepackage{braket}
\begin{document}
%\dr
\title{ Microstructure from  ferroelastic transitions using strain pseudospin clock models 
in two and three dimensions: a local mean-field analysis }
\author{ Romain Vasseur$^{1,2}$}
\author{Turab Lookman$^{1}$}
\author{Subodh R. Shenoy$^{3}$}

\affiliation{$^1$Theoretical Division and Center for Nonlinear Studies, Los Alamos National Laboratory,
Los Alamos, New Mexico 87545, USA \\
$^2$ \'Ecole Normale Sup\'erieure de Lyon, 46 All\'ee d'Italie
69007 Lyon, France \\
$^3$ School of Physics, University of Hyderabad, Hyderabad 500046, India }

\begin{abstract}
 
 We show how microstructure can arise in first-order ferroelastic structural transitions, in two and three spatial dimensions, through a local meanfield approximation of their pseudospin hamiltonians, that include  anisotropic elastic interactions. Such  transitions  have symmetry-selected physical strains as their  $N_{OP}$-component  order parameters, with  Landau free energies that have   a  single zero-strain 'austenite' minimum at high temperatures,  and  spontaneous-strain  'martensite' minima  of  $N_V$ structural variants at low temperatures.  The total free energy also has gradient terms, and powerlaw anisotropic effective interactions, induced by 'no-dislocation' St Venant compatibility constraints. In a reduced description, the strains at Landau minima   induce temperature-dependent, clock-like $\mathbb{Z}_{N_V +1}$  hamiltonians, with  $N_{OP}$-component strain-pseudospin vectors  ${\vec  S}$  pointing to  $N_V + 1$ discrete values (including zero). We study elastic texturing in five such first-order structural transitions through a local meanfield approximation of their pseudospin hamiltonians, that include the powerlaw interactions. As a prototype, we consider  the two-variant square/rectangle transition, with a one-component, pseudospin   taking  $N_V +1 =3$ values of $S= 0, \pm 1$,  as  in a generalized Blume-Capel model. We then  consider  transitions with two-component ($N_{OP} = 2$) pseudospins:   the  equilateral to centred-rectangle ($N_V =3$); the  square to oblique polygon ($N_V =4$); the triangle to oblique  ($N_V =6$) transitions; and finally  the 3D  cubic to tetragonal transition ($ N_V =3$).  The local meanfield solutions   in 2D and 3D yield oriented domain-walls patterns  as from continuous-variable strain dynamics, showing  the discrete-variable models capture the essential ferroelastic texturings. Other related hamiltonians illustrate that structural-transitions in  materials science can be the source  of interesting spin models in statistical mechanics.
\end{abstract}


\maketitle

\section{Introduction}
  Ferroelastic crystals undergo diffusionless structural transitions that are  first order, and on cooling show  a reduction in symmetry to two or more spontaneously-strained states (or 'variants') which can be transformed between one  another by stress~\cite{R1,R2}. These transitions  are often studied through minimizing the Landau free energies~\cite{R3} in terms of 
 appropriate continuous  variables, such as displacements, phase fields, or  strains~\cite{R4,R5,R6,R7}. Although homogeneous, single-variant martensite states are the  global minimum,  elastic heterogeneities or metastable domain-wall patterns are experimentally found \cite{R8},  that are  locked in to preferred crystallographic directions~\cite{R9}. This orientation arises through a balance between the Landau energies nonlinear in the order parameter strain, the short-range gradient costs or Ginzburg energies, and the effectively long-range elastic energies \cite{R10} or  powerlaw anisotropic interactions, that orient the domain walls ~\cite{R5,R7,R10}. The powerlaw  interactions result from enforcing St Venant 'compatibility' constraints~\cite{R11,R12} between the strain components
so that the displacements are continuous, with no dislocations generated on cooling.  Continuous-variable models have been used to study microstructures under various  conditions, including strain-rate dependence~\cite{R13}; and  the effects of finite-size on martensitic growth in an austenitic matrix ~\cite{R14}.

Models in terms of {\it discrete} structure-variables or 'pseudospins'  have also been used to study  these ferroelastic transitions~\cite{R15,R16}. (Similar in spirit to discrete strains, analytic `minimizing sequences'  consider tent-like displacement profiles or flat strain-variants, on either sides of domain walls  \cite{R2}.) Recently,  model pseudospin hamiltonians induced by the scaled free energies  for several  specific transitions in 2D and 3D have been proposed~\cite{R17}. The model hamiltonian   is simply  the total  scaled~\cite{R18} free energy evaluated at the Landau minima in the order parameters (OP). The pseudospins are 'arrows' in $N_{OP}$-dimensional order-parameter space, pointing to the $N_V$ variant minima, and to the zero-strain turning point. The hamiltonian includes a temperature-dependent on-site  term quadratic in the pseudospins from the Landau term, a nearest-neighbor ferromagnetic interaction between pseudospins from the Ginzburg term, and a pseudospin powerlaw interaction from the St Venant term. The pseudo-spin models are like $\mathbb{Z}_{N_V +1}$ clock models~\cite{R19} generalized to  include a spin-zero state,  and may be termed  'clock-zero' models~\cite{R17}. A three-state spin-1 type model for the transition of square to rectangle unit cells (with $N_{OP} =1, N_V =2$)  has found glass-like behavior on slow cooling using a local meanfield approximation~\cite{R17}. 

In this work  we consider  pseudo-spin models in the {\it  local meanfield approximation} under temperature quenches, for five structural transitions : four  in two spatial dimensions~\cite{R20}, and one  in three spatial dimensions~\cite{R7}.  Apart from the single order parameter  ($N_{OP} =1$) square/rectangle case that is first studied as a simple prototype, the other four transitions all have two-component ($N_{OP} = 2$) pseudospins. The transitions  are i) the square to rectangle (2D version of the  tetragonal/orthorhombic transition such as in YBCO);  ii) the square to oblique polygon;  iii) the triangle to centred-rectangle (2D version of hexagonal to orthorhombic transition such as in lead orthovanadate~\cite{R4,R6,R7}); iv)   the triangle to oblique;  and v)  in 3D, the cubic to tetragonal transition (as in FePd). For these  five transitions,  the (nonzero) pseudospin arrows point respectively, to the two ends of a line, and to the corners of a square, a triangle, a hexagon, and a triangle, with the number of pseudospin variant states thus being respectively $N_V = 2, 4, 3, 6$ and $3$.   We show that these discrete-variable models, despite their simplicity, produce  local-meanfield microstructure  in one-  and two-component strain pseudospins in agreement with continuous-variable strain simulations~\cite{R4,R5,R6,R7},  that can be computationally more intensive.  We thus find parallel  twins for the square/rectangle and cubic/tetragonal transitions; nested stars for the equilateral/isosceles triangle transition; and tilted oblique domains for the square/oblique, and triangle/oblique transitions.

The generalized clock model  of strain pseudospins is a statistical mechanics description of the ferroelastic transitions  in materials science. It conceptually links long-lived, metastable martensitic twins (even without quenched disorder)  to   Potts-model  and clock-model descriptions of glasses~\cite{R19}, and may be relevant  to  recent quenched-disorder strain glass behavior  in martensitic alloys~\cite{R21}.

The plan of the paper is as follows. In Section~\ref{Section2} we outline the derivation~\cite{R17} of the pseudospin hamiltonians, and of compatibility potentials for the four 2D transitions. Our  meanfield microstructure results are in Section~\ref{Section3}  where  we first consider  the two-variant square/rectangle case as a prototype, its response to external stress, and  its relation to the  spin-1 Blume-Capel model~\cite{R21}. We then consider local meanfield microstructure for the three-variant  triangle to centred-rectangle transition; the  four-variant square/oblique transition; and   the six-variant  triangle/oblique transition. Turning to 3D, Section~\ref{Section4} considers the local meanfield microstructure for the three-variant  cubic/tetragonal transition, with its compatibility potential  stated in the Appendix. In Section~\ref{Section5} we mention other related spin models  of interest in statistical mechanics.    The final Section ~\ref{Section6} has a summary and conclusion.

\section{PSEUDOSPIN HAMILTONIANS IN TWO SPATIAL DIMENSIONS}

\label{Section2}

The free energy functionals describing ferroelastic structural transitions can be written  in terms of the {\it physical } strains, that are symmetry-specific linear combinations of the Cartesian strain-tensor components. The Landau terms are invariant polynomials of the $N_{OP}$ order parameter strains, and have $N_V$    minima.   The free energies have many material-dependent  elastic coefficients, that are not always known, or are fitted to experiment only for specific materials.   However, the spontaneous order-parameter strain magnitude at the first-order transition temperature is a small parameter. Following Barsch and Krumhansl~\cite{R18} a scaling procedure has been applied~\cite{R17} to four 3D transitions and five 2D transitions to obtain scaled Landau free energies that (to leading order in the small parameter)  show  universality at their  minima, where any  internal elastic constants are scaled out, and  material dependence is only in an overall elastic-energy prefactor.  `Geometric nonlinearities` are higher order in the spontaneous strain and are  neglected,  as a perturbative first approximation. Then  different materials with  the same transition, fall into  the same 'quasi-universality'  class, with common  behaviour at the scaled minima, that lie  at the corners and centres of  the same 'polyhedron' in $N_{OP}$ dimensional order parameter space. This is useful in strain-variable dynamics. It also immediately suggests a reduced description of ferroelastics, in terms of discrete-strain statistical variables or vector 'pseudospins', directed to these minima.

%The general  approach  of using discrete-variable pseudospins to approximate continuum-variable  distortions has been  pursued in various models ~\cite{R15,R16,R17,R20}.
 A specific reduction procedure  was proposed  \cite{R17} to obtain pseudospin hamiltonians by evaluating  {\it scaled}  free energies evaluated at their Landau minima.
The basic idea  is quite simple. i) Scale the total free energy to dimensionless form, including the specifically calculated compatibility-induced powerlaw interaction term, and the gradient term. Write the Landau free energy  in polar coordinates in OP space, with  the austenite minimum at the origin, and   $N_V$ martensite minima located on a circle in  $N_V$ discrete angular directions.  ii) Set the  radial OP-magnitude to its common temperature-dependent Landau-minimum value, and replace the OP-minima directional angles by discrete vectors  pointing to these  $N_V + 1$ minima on the circle, and at the centre. \\
 iii) The total free energy evaluated at minima  is then the model Hamiltonian for the vector pseudospins, that have $N_{OP}$ spin-components, and $N_V + 1$ values. The remaining model coefficients are then not just arbitrary, but are related  through the parent free energy,  to the scaled temperature, to the scaled energy cost of an elastic domain-wall segment, and to the scaled bulk stiffness. 

We outline below the derivation~\cite{R17} of the pseudospin hamiltonians and compatibility potentials in two spatial dimensions, for the square/rectangle, triangle/ /centred-rectangle, square/oblique, and triangle/oblique transitions, with number of variants  $N_V = 2, 3, 4$ and $6$ respectively. The 3D case is considered later.

 \subsection{Square to Rectangle (SR) hamiltonian: $N_{OP} =1, N_V =2$}         
 
Consider the prototypical square-to-rectangle or \mbox{'SR'}  transition, that is a two-dimensional analog of a  tetragonal to orthorhombic transition. 
For small distortions, the components of the symmetric Cartesian strain tensor are given by $e_{\mu \nu} = 1/2 \ ( \partial_\mu u_\nu +
   \partial_\nu u_\mu )$, where $ \vec{u} (\vec r)$ is the displacement vector and $\mu , \nu = x, y$.  We define linear combinations of the Cartesian components as three  physical  strains, describing compressional ($e_1$), deviatoric ($e_2$) and shear ($e_3$) distortions.
\begin{equation}
e_1=\frac{c_1}{2} \ (e_{x x}+ e_{y y}), \ \
e_2=\frac{c_2}{2} \ (e_{x x}- e_{y y}), \ \
e_3= \frac{c_3}{2} (e_{x y} + e_{yx}),
\label{2.1}
\end{equation}
where $c_1$, $c_2$ and $c_3$ are symmetry-specific constants~\cite{R7}. For the square reference lattice, $c_1 = c_2 = \sqrt{2}$ and $c_3 = 1$.  For the equilateral triangle reference  lattice $c_1 = c_2 = c_3 =1$.
The pseudospin hamiltonian is obtained by the three steps given above, that we follow for all transitions.

{\it  i) Scaled free energy, and  compatibility potential :}

For the SR case, the deviatoric strain $e_2$  is the order parameter (OP).  The compressional and shear strains are the non-OP  strains.  The scaled free energy is $F = E_0  {\bar F}$ where  the overall $E_0$ is an  elastic energy per unit cell, and the dimensionless $\bar F= \bar{F}_L + \bar{F}_G  +\bar{F}_{non} $ is a sum of three terms, 
\begin{equation}
\bar{F}_L = \sum_{\vec r}{\bar f}_L (e_2); ~ \bar{F}_G = \sum_{\vec r} {\bar f}_G (\vec \nabla e_2); ~  \bar{F}_{non} = \sum_{\vec r} {\bar  f}_{non} (e_1, e_3). 
\label{2.2}
\end{equation}
where $\sum_{\vec r} \rightarrow \int d^2 r /a_0^2$ runs over all positions, and $a_0$ is a lattice scale for a computational grid. 

The dimensionless, scaled Landau free energy density in coordinate space  is sixth order in $e_2 (\vec r)$ to give  a first order transition~\cite{R5,R7}, 

\begin{equation}
\displaystyle{\bar f}_{L} (e_2) =  (\tau-1) e_2^2 + e_2^2 (e_2^2-1)^2  .
\label{2.3}
\end{equation}
A scaled temperature is defined by
\begin{equation}
\displaystyle{\tau \equiv ( T - T_c)/(T_0 - T_c).}
\label{2.4}
\end{equation}

 There can be three  $\partial f_L /\partial e_2 =0$ minima:  at zero-strain $e_2 =0$ austenite, and at two martensite variant minima of nonzero strain, $e_2 = \pm \bar{\varepsilon} (\tau)$. The order-parameter magnitude $\bar \varepsilon$ at the variant minima is   
\begin{equation}
\displaystyle{ {\bar \varepsilon}(\tau)= [ \frac{2}{3} ( 1 + \sqrt{1 - 3\tau / 4})]^{1/2} .}
\label{2.5}
\end{equation}
On cooling below the upper spinodal $\tau =4/3$, two martensite variants  appear ;  they become  degenerate with the austenite zero-state at $\tau = 1$ or $T= T_0$ when ${\bar \varepsilon}(\tau =1) = 1$; and  for $\tau < 1$ the martensite wells become lower in energy. The austenite minimum disappears below the lower spinodal $\tau= 0$ or $T =T_c$. 

The cost of creating interfaces or domain walls is given by the usual Ginzburg term, with $\xi$ a wall thickness scale, 
\begin{equation}
\displaystyle{ \bar{F}_{G} = \sum_{\vec r}  \xi^2  ( {\vec \nabla} e_2 (\vec r) )^2 = \sum_{\vec k}  \xi^2  {\vec k} ^2 | e_2 (\vec k) |^2. }
\label{2.6}
\end{equation}

Finally, the non-OP strain  energy is simply harmonic in  compressional ($e_1$)  and shear ($e_3$) strains, 

\begin{equation}
\displaystyle{\bar{F}_{non}  = \sum_{{\vec r} , i =1,3} \frac{1}{2}  A_i {e_i (\vec r)}^2 = \sum_{{\vec k} , i =1,3} \frac{1}{2} A_i |e_i (\vec k)|^2 . }  
\label{2.7}
\end{equation} 
The scaled compressional and shear elastic constants  can be expressed in terms of the (unscaled) elastic constants $C_{i j}$  in the Voigt notation, evaluated at $T_0$.  For the cubic case~\cite{R17} , $A_1= (C_{11}  + 2 C_{12}) / ( C_{11}  - C_{12})$.   The ratio $A_1 /A_3$ is taken as fixed in simulations, for simplicity.

For $\vec k =0$ uniform contributions, the optimum non-OP strains are zero, at the parabolic $f_{non} =0$ minimum. For spatially varying $\vec k \neq 0$ contributions,
the non-OP strains are to be minimized subject to the St Venant compatibility constraint~\cite{R5,R7,R11,R12} that says distorted unit cells fit together in a smoothly compatible fashion, without defects like dislocations, so the displacement field is single-valued. The St. Venant conditions in the Cartesian strain tensor ${\bf e}$ are ~\cite{R11,R12} (with ' T ' a transpose),
\begin{equation}
\displaystyle{{\vec  \nabla} \times ({\vec \nabla} \times { \bf e } ) ^ T = 0 .}
\label{2.8}
\end{equation}

In two dimensions, the constraint  in terms of physical strains of~\eqref{2.1} is 
\begin{equation}
\displaystyle{ \frac{1}{c_1} {\vec \nabla} ^2 e_1 - \frac{1}{c_2}(\partial^2_{x} - \partial^2_{y}) e_2 - \frac{2}{c_3} \partial_{x} \partial_{y} e_3 = 0 .}
\label{2.9}
\end{equation}  
or in Fourier space

\begin{equation}
\displaystyle{  O_1 e_1(\vec k) +  O_2 e_2(\vec k)   +O_3 e_3(\vec k)  =0 ,}
\label{2.10}
\end{equation}  
where the compatibility coefficients are
\begin{equation}
\displaystyle{O_1=\frac{- \vec{k}^2}{c_1}}, \ \
\displaystyle{O_2=\frac{k_x^2-k_y^2}{c_2}}, \ \
\displaystyle{O_3=\frac{2 k_x k_y}{c_3}} .
\label{2.11}
\end{equation} 
The constrained minimization can be done through Lagrange multipliers~\cite{R5}, or by a direct substitution of the constrained solution~\cite{R17}  $e_1 = - (O_2 e_2 + O_3 e_3)/O_1$  of~\eqref{2.10}, into the non-OP free energy of~\eqref{2.7},
\begin{equation}
\displaystyle{\bar{f}_{non}= \sum_{\vec k \neq 0} \frac{1}{2} \left[ A_1~ | (O_2 e_2 + O_3 e_3)/O_1|^2 + A_3 |e_3|^2 \right] . }
\label{2.12}
\end{equation}
A free minimization in the remaining  non-OP strain $e_3$ determines it in terms of the OP $e_2$. In fact, $e_i = - (O_i O_2 /A_i)/[(O_1 ^2/A_1) + (O_3 ^2/A_3)]$ for $i = 1, 3$. Substituting into~\eqref{2.7} yields the compatibility-induced interaction $\bar{F}_{compat} (e_2)  \equiv \bar{F}_{non} (e_1 , e_3)$, 
where
 \begin{equation}
\displaystyle{\bar{F}_{compat} = \sum_{\vec k} \frac{A_1}{2}U(\vec k) |e_2 (\vec k)|^2 }. 
\label{2.13}
\end{equation}
The compatibility kernel $U$ in Fourier space  is
\begin{equation}
\displaystyle{ A_1 U(\vec k) = \nu(\vec k) \dfrac{O_2 ^2}{[(O_1 ^2/A_1) + (O_3 ^2/A_3)]} },
\label{2.14}
\end{equation} 
or explicitly from (11), and $c_1 = c_2 = \sqrt{2} , c_3 = 1$, 
\begin{equation}
\displaystyle{U(\vec k) = \nu(\vec k) \dfrac{(k_x^2-k_y^2)^2}{k^4+8\frac{A_1}{A_3}k_x^2 k_y^2  } }.
\label{2.15}
\end{equation} 
Here the prefactor $\nu (\vec k) \equiv 1 - \delta_{{\vec k}, 0}$ is inserted to make $f_{non} \sim U$ vanish for $\vec k =0$ uniform non-OP strains, as mentioned. In coordinate space, this is a powerlaw interaction between OP strains $U(\vec R) \sim 1/R^d$, with sign-variation in  angular directions yielding zero angular average ($\sum_{\vec R} U(\vec R) \sim U(\vec k =0) =0$), so it is not 'long-range' in the isotropic Coulomb $\sim 1/R^{d-2}$ sense. The powerlaw anisotropic interactions are easily evaluated in Fourier space, and one need not resort to uncontrolled coordinate-space  truncations to near-neighbor couplings, that may leave out some essential physics of the transition. 
In coordinate space,

\begin{equation}
\displaystyle{\bar{F}_{compat}  (e_2 )= \frac{1}{2}  \sum_{{\vec r} ,{ \vec r'}}  A_1 U(\vec{r}-\vec{r}') e_2(\vec{r}) e_2(\vec{r}')}.
\label{2.16}
\end{equation}

The formal partition function 

\begin{equation}
\displaystyle{ Z = \int \prod_{\vec{r}} d e_2(\vec{r}) \exp(- \beta F[e_2( \vec{r} )])},
 \label{2.17}
\end{equation} 

\noindent  is dominated by free energy textural minima, that may be  asymptotically found in a TDGL or relaxational dynamics, as done elsewhere~\cite{R7},
\begin{equation}
\displaystyle{ \frac{\partial e_2(\vec r, t)}{\partial t} =  - \frac{\delta F}{\delta e_2 (\vec r, t)}}.
\label{2.18}
\end{equation}

 {\it ii) Continuous strains to discrete pseudospins :}
 
 One can approximate  the partition function by  retaining only  the Landau-minima at fixed OP-magnitude values $|e_2| =\bar{\varepsilon} (\tau)$,  and  different OP signs (or in general, different angular directions of minima), while neglecting fluctuations about these minima.  
The continuous-variable strains are then replaced by discrete-variable pseudospins~\cite{R17}

\begin{equation}
\displaystyle{e_2(\vec{r}) \rightarrow {\bar \varepsilon} (\tau) S(\vec{r})},
\label{2.19}
\end{equation}
 where the pseudospin  has the three values $S(\vec{r})=0, \pm 1$, to locate the minima at $e_2 = 0, \pm {\bar \varepsilon} (\tau)$. Although in zero stress  the uniform austenite state is no longer a Landau minimum  below the lower spinodal $\tau = 0$, the surrounding nonuniform textures  can exert local internal stresses to locally favor the zero value, even at  low temperatures.  Also, the free energy in OP strain always has a turning point at the origin to support dynamical transient zeros, that although few in number, could play a catalytic role in microstructural evolution~\cite{R17}. Hence we retain zero spin values at all temperatures, allowing their permanent/transient existence to be determined dynamically. 

With this substitution and $S^6 = S^4 = S^2 = 1, 0$, the approximated Landau free energy density at the minima  can be written as~\cite{R17}

\begin{equation}
\displaystyle{{\bar f}_L = {\bar \varepsilon}^2 (\tau) g_L (\tau) S^2 (\vec r) ; ~~g_L \equiv  (\tau - 1) +  ({\bar \varepsilon}^2 -1)^2.}
\label{2.20}
\end{equation}
where $\bar \varepsilon$ is in~\eqref{2.5}.

{\it  iii) The reduced  pseudospin hamiltonian :}

The  partition function of~\eqref{2.17} reduces to a sum over all the pseudospin  configurations, with a  temperature-dependent effective Hamiltonian in the Boltzmann weight, that  can then be studied by the usual methods of statistical mechanics.
Substituting~\eqref{2.19} into the total scaled free energy directly yields the hamiltonian in coordinate space,

\begin{equation}
\displaystyle{ H(S) \equiv    {\bar F}(e_2 \rightarrow \bar{\varepsilon} S) },
\label{2.21}
\end{equation}
where

\begin{align}
\displaystyle \beta H(S) & = \frac{D_0}{2}[ \sum_{\vec r}  \{g_L S^2 (\vec r) + \xi^2 ( \vec{\nabla} S)^2\}  \\ \notag
\displaystyle  & +  \sum_{{\vec r}, {\vec r'}} \frac{A_1}{2} U({\vec r} - {\vec r'}) S(\vec r) S(\vec r') ],
\label{2.22}
\end{align}

\noindent and $D_0 \equiv 2 E_0 \bar{\varepsilon} ( \tau)^2/T$.  
 This has the form of a generalized spin-1  Blume-Capel model~\cite{R22} as discussed later, but with  temperature-dependent  coefficients  and  powerlaw interactions.
    The hamiltonian  is diagonal in Fourier space \cite{R17},
\begin{equation}
\displaystyle{\beta H =  \frac{1}{2} \sum_{\vec k}   Q_0 (\vec k)|S (\vec k)|^2},
\label{2.23}
\end{equation}
where
\begin{equation}
\displaystyle{Q_0  (\vec k) \equiv D_0 [ g_L (\tau) +  \xi^2 {\vec k}^2   + \frac{A_1}{2} U (\vec k )] }.
\label{2.24}
\end{equation}

\subsection{Triangle/Centred Rectangle  (TCR) hamiltonian:  $N_{OP} =2, N_V =3$}

Consider a two-dimensional  crystal with equilateral triangles transforming to isosceles triangles, with three possible such variants ($N_V = 3$), as there are three sides that can become the unequal side. The unit-cell changes from an equilateral triangle to a  centred-rectangle.
This 'TCR'  transition is the 2D version of the hexagonal to orthorhombic transition
observed in lead orthovanadate~\cite{R8}.
There are two order parameters~\cite{R4,R7,R17,R20}: the deviatoric strain  $e_2$, and the shear strain $e_3$. The single non-OP variable  is the bulk dilatation or compressional strain $e_1$.  Just like this TCR case, the square/oblique, triangle/oblique and cubic/tetragonal   transitions also have the same OP $(e_2 , e_3)$  and  a single non-OP strain $e_1$, but of course are distinguished by their different, transition-specific Landau polynomials, that induce different $N_V$ directions of the vector pseudospins. 

{\it i) Scaled free energy and compatibility potential :}

 The free energy functional, invariant under the triangular point group symmetry, is  

\begin{equation}
\displaystyle{ \bar{F} = \sum_{\vec r} \left\lbrace  {\bar f}_{L} \left(e_2, e_3 \right) + {\bar  f}_{G} \left({\vec \nabla} e_2, {\vec \nabla} e_3 \right) + {\bar f}_{non} \left(e_1\right) \right\rbrace }.
\label{2.25}
\end{equation}

\begin{figure}[h]
\includegraphics[width=8.0cm]{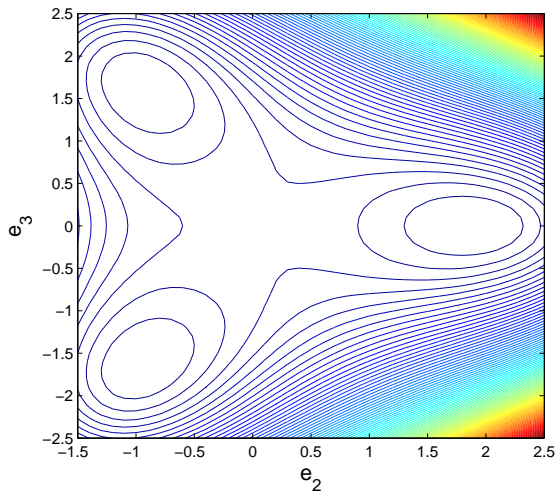}
\caption{Contour plot  in $(e_2 , e_3)$ space of the Landau free energy for the Triangle to  Centred-Rectangle (TCR) transition, with  parameters $\tau = - 2.5$ and $E_0 = 1$. The three
degenerate energy minima correspond to the three martensite variants at this low temperature.} 
\label{TR_landau}
\end{figure}

The Landau free energy $f_{L}$ for the TCR case  describes the first-order phase transition between the single high-symmetry austenite phase and the $N_V = 3$  martensite variants. It has a third-order  term invariant under equilateral triangle symmetries, $I_3 \equiv e_2^3-3 e_2 e_3^2 $. In scaled form, in coordinate space

\begin{equation}
\displaystyle{{\bar f}_{L} (e_2, e_3) =   \tau (e_2^2+e_3^2) - 2 (e_2^3-3 e_2 e_3^2) + (e_2^2+e_3^2)^2.}
\label{2.26}
\end{equation}
 \mbox{Figure \ref{TR_landau}} shows the Landau free energy with three variant minima,  for a low temperature.
 
In polar coordinates in OP space, the  order parameter vector  is $\vec e (\vec r)  = (e_2, e_3) = \varepsilon (\cos \phi, \sin \phi)$ with  magnitude $\varepsilon (\vec r)  \equiv (e_2 ^2 +e_3 ^2)^{1/2}$. The 
Landau free energy in polar coordinates, with $\eta_3 \equiv \cos 3 \phi$  is~\cite{R17}
\begin{equation}
\displaystyle{ {\bar f}_L (\varepsilon, \phi) = [ (\tau -1) \varepsilon^2 + \varepsilon^2(\varepsilon-1)^2] +  2(1-\eta_3)\varepsilon^3}
\label{2.27}
\end{equation}

The angular dependence is  ${\bar f}_L \sim - \cos 3 \phi$. The minimum conditions  $\partial f_L /\partial \varepsilon =0; \partial f_L /\partial \phi =0$ yield four minima: at the $\vec e =0$ austenite, and the three  variant minima with $\sin 3 \phi =0$  where $\phi =  \phi_m = 0, \pi/3, 2 \pi /3$, so   the last term  in~\eqref{2.27} vanishes.  
 The three variant minima in the $(e_2 , e_3)$ plane form a triangle lying on a circle of radius $\varepsilon = \bar{\varepsilon} (\tau)$, where
\begin{equation}
\displaystyle{ \bar{\varepsilon} (\tau)= \frac{3}{4} ( 1 + \sqrt{1 - 8 \tau / 9})  }.
\label{2.28}
\end{equation}

On cooling below the upper spinodal $\tau =4/3$, two martensite variants  appear ;  they become  degenerate with the austenite zero-state at $\tau = 1$ or $T= T_0$ when ${\bar \varepsilon}(\tau =1) = 1$; and  for $\tau < 1$ the martensite wells become lower in energy. The austenite minimum disappears below the lower spinodal $\tau= 0$ or $T =T_c$.

The Ginzburg term ${\bar f}_{G}$ is quadratic in the OP-strain gradient  so 

\begin{equation}
\displaystyle{ {\bar{F}}_{G} =  \sum_{{\vec r}, \ell =2,3} \xi^2 ({\vec \nabla} e_\ell) ^2  =\sum_{{\vec k}, \ell =2,3} \xi^2 {\vec k} ^2 |e_\ell (\vec k) |^2 }.
\label{2.29}
\end{equation}

\paragraph*{}
Finally, the non-OP term is simply harmonic in the single non-OP compressional strain,
\begin{equation}
\displaystyle{  \bar{F}_{non} (e_1) =\frac{A_1}{2}  \sum_{\vec r} e_1^2 (\vec r)=  \frac{A_1}{2}  \sum_{\vec k} |e_1 (\vec k)|^2 }.
\label{2.30}
\end{equation}
 
 Substitution of the compatibility solution $e_1 (\vec k) = - (O_2 e_2  + O_3 e_3) / O_1$  as for  (12) immediately yields the St Venant term                                                                                      $\bar{F}_{compat} (e_2, e_3) \equiv  {\bar F}_{non} (e_1)$ in terms of the OP,
 
 \begin{equation}
\displaystyle{ \bar{F}_{compat} = \sum_{\vec k , \ell, \ell' = 2,3}  \frac{A_1}{2} U_{\ell \ell'} (\vec k) e_\ell (\vec k) e_{\ell' }(\vec k)^*} .
\label{2.31}
\end{equation}
The compatibility potential kernel is~\cite{R7,R15}
\begin{equation}
\displaystyle{U_{\ell \ell'} = \nu(\vec k) O_{\ell} O_{\ell'}/O_{1}^2}, 
\label{2.32}
\end{equation}

\noindent or explicitly from~\eqref{2.11} with $c_1 = c_2 = c_3 =1$,

\begin{align}
\displaystyle & U_{22}  = \nu \frac{(k_x ^2 - k_y ^2)^2}{k^4}; ~ U_{22} = \nu \frac{(2 k_x  k_y )^2}{k^4}; \notag \\ 
\displaystyle & U_{23} = \nu \frac{2 k_x k_y (k_x ^2 - k_y ^2)}{k^4} = U_{32} .
\label{2.33}
\end{align}

In coordinate space, 
\begin{equation}
\displaystyle{ \bar{F}_{compat}  (e_2, e_3 )= \frac{A_1}{2}\sum_{\vec r, \vec r'}  \sum_{\ell,\ell' = 2,3}   U_{\ell \ell'}(\vec{r}-\vec{r}') e_\ell(\vec{r}) e_{\ell'}(\vec{r}')},
\label{2.34}
\end{equation}
and as before, the powerlaw potentials fall off in 2D as $U_{\ell \ell'} \sim 1/ R^2$.

{\it ii) Continuous strains to discrete  pseudospins :}

For the TCR case (and  other two-component OP  cases),  $N_V \geq  3$, but  we {\it do not} simply  get a generalized spin-$j$ model with $2j +1$ states on a line,  and $j = N_V/2$. Instead we obtain {\it  clock-like}  models~\cite{R17,R19} with discrete $\vec{S}$ vector variables pointing to the polyhedron$N_V$ corners and centre,  in  $N_{OP}$-dimensional space. Since the zero state is included, these  may be termed 'clock-zero'  ${\mathbb{Z}}_{N_V +1}$ models~\cite{R17}. Note that, unlike pure clock ${\mathbb{Z}}_N$ models~\cite{R17}, the squared spin         $\vec{S}(\vec r)^2 =1,0$ is still a statistical variable and not a constant, because of the zero states.  

The continuous-variable strains at minima are replaced by discrete-valued pseudospins~\cite{R17},

\begin{equation}
\displaystyle{\vec{e} (\vec{r})= \begin{pmatrix} e_2 \\ e_3 \end{pmatrix} \longrightarrow \bar{\varepsilon} (\tau)\begin{pmatrix} S_2 (\vec r) \\ S_3 (\vec r) \end{pmatrix} },
\label{2.35}
\end{equation}

where the two components of the pseudospin have three variant values as in Fig~\ref{TR_landau} plus zero, 
\begin{equation}
\displaystyle{\vec{S} =  \begin{pmatrix} 0 \\ 0 \end{pmatrix},  \begin{pmatrix} 1 \\ 0 \end{pmatrix},  \begin{pmatrix} -\frac{1}{2} \\ \pm \frac{\sqrt{3}}{2} \end{pmatrix} }.
\label{2.36}
\end{equation}
For the $N_V =3$ variants, $\vec S = (\cos \phi_m, \sin \phi_m)$ and ${\vec S}^2 =1$, with $\phi_m = 2 \pi (m-1)/ 3$ and $m = 1, 2, 3$.

With this substitution and ${\vec S} ^6 = {\vec S} ^4 = {\vec S} ^2= 0, 1$, the Landau polynomials again collapse  into a simple form, bilinear in the pseudospins,
\begin{equation}
\displaystyle{{\bar f}_L  (\tau) = \bar{\varepsilon} ^2  g_L {\vec S} ^2; ~~ g_L (\tau)  \equiv  \tau - 1 +( \bar{\varepsilon} -1)^2},
\label{2.37}
\end{equation}
with $\bar \varepsilon$ as in~\eqref{2.28}.

{\it iii) The reduced pseudospin hamiltonian : }

In coordinate space the total pseudospin hamiltonian is

\begin{equation}
\begin{array}{rr}
\beta H & = \dfrac{D_0}{2} [ \sum_{\vec{r}} \sum_{\ell = 2,3} \{g_L {S_\ell}(\vec{r})^2 + \xi^2 |\vec \nabla S_\ell(\vec{r})| ^2\} \\\ & 
+ \dfrac{ A_1}{2} \sum_{\vec{r},\vec{r}'}\sum_{\ell,\ell' =2, 3} U_{\ell\ell'}(\vec{r}-\vec{r}') S_{\ell}(\vec{r}) S_{\ell'}(\vec{r}' )] ,
\end{array}
\label{2.38}
\end{equation}  
and is a clock-zero $\mathbb{Z}_{3+1}$ model~\cite{R17} , with  $\vec S$ having $3 + 1$ values of~\eqref{2.36}, and with a compatibility kernel  of~\eqref{2.33}. 
It is again diagonal in Fourier space,
\begin{equation}
\displaystyle{\beta H = \frac{1}{2} \sum_{\vec k} \sum_{ \ell, \ell'}  Q_{0, \ell \ell'} (\vec k) S_\ell (\vec k) S_{\ell'} (\vec k)^* },
\label{2.39}
\end{equation}
with  $\vec{S}(\vec{k})^{*} = \vec{S}(- \vec{k})$, as ${\vec S} (\vec r)$ is  real. Here
 \begin{equation}
\displaystyle{Q_{0, \ell \ell'} (\vec k) \equiv D_0 [\{ g_L(\tau) +  \xi^2 { \vec k}^2  \} \delta_{\ell, \ell'} + \frac{A_1}{2} U_{\ell  \ell'} ({\vec k} )].}
\label{2.40}
\end{equation}

\subsection{Square/Oblique (SO)  hamiltonian: $N_{OP}=2, N_V =4$}
 
 We consider the square/oblique  or ' SO ' transition where  the transition is driven independently by the deviatoric $e_2$ and shear $e_3$ order parameter strains~\cite{R7,R17,R20}, as modified by a sufficiently strong coupling term.

 {\it i) Scaled free energy, and  compatibility potential :}
 
  The Landau term has the scaled form 

\begin{equation}
\displaystyle{ {\bar f}_{L} =   \tau (e_2^2+e_3^2) - (4 - C' _4 /2) (e_2^4 + e_3^4) + 4 (e_2^6+e_3^6) - C'_4 e_2^2 e_3^2 },
\label{2.41}
\end{equation}
where $C'_4$ is a material-dependent elastic constant. In polar coordinates,  with $\vec e = (e_2 , e_3) = \varepsilon (\cos \phi, \sin \phi )$, it is~\cite{R17}

\begin{equation}
\displaystyle{ \bar{f} _{L}=[(\tau-1)\varepsilon^2+\varepsilon^2(\varepsilon^2-1)^2] +  \varepsilon^4 (3\varepsilon^2- 2 + C'_4 /2 )\cos^2 2 \phi }.
\label{2.42}
\end{equation}
The angular dependence is ${\bar f}_L \sim \cos 4 \phi$. The five minima from $\partial {\bar f}_L /\partial \varepsilon =0; \partial {\bar f}_L/\partial \phi =0$ are the austenite zero state, and four variant minima with $\sin 4 \phi =0$  in 
 angular directions $\phi = \phi_m = \pi (2m -1) /4$ with $m = 1, 2, 3, 4$. The last term  in~\eqref{2.42} vanishes at minima, suppressing the $C_4 '$ material dependence.
The four  variant minima in the $e_2, e_3$ plane for $\tau < 4/3$ form a square lying on a circle of radius $\varepsilon = \bar{\varepsilon} (\tau)$, where $ \bar{\varepsilon}$ is as in  the SR case of~\eqref{2.5}.
%\begin{equation}
%\displaystyle{ \bar{\varepsilon} (\tau)= {\frac{2}{3} ( 1 + \sqrt{1 - 3\tau / 4})^{1/2} }.
%\label{2.42}
%\end{equation}
%as in the SR case. 

{\it ii) Continuous strains and discrete pseudospins :}
The strains at minima are replaced by  pseudospins as in~\eqref{2.35}.
The discrete pseudo-spin has the five values~\cite{R17}

\begin{equation}
\vec{S} =  \begin{pmatrix} 0 \\ 0 \end{pmatrix},  \begin{pmatrix} 1 \\ 0 \end{pmatrix},  \begin{pmatrix} \pm\frac{1}{\sqrt 2} \\ \pm \frac{1}{\sqrt 2}  \end{pmatrix}.
\label{2.43}
\end{equation}
For the $N_V = 4$ variant minima with $\vec S = (\cos \phi_m , \sin \phi_m )$, and $\phi = \phi_m = \pi(2 m -1) /4$ where $m =1, 2,3,4$ , the spin magnitude is unity ${\vec S}^2 =1$.

The Landau term becomes
\begin{equation}
\displaystyle{\bar{f}_L = \bar{\varepsilon} ^2  g_L {\vec S} ^2; ~~ g_L \equiv  \tau - 1 +( \bar{\varepsilon}^2 -1)^2}
\label{2.44}
\end{equation}
with $\bar \varepsilon$ of (5). 

 {\it iii) The reduced pseudospin hamiltonian: }

The Ginzburg and St Venant terms are the same as in the TCR case.
The SO case clock-zero  hamiltonian ${\mathbb{Z}}_{4 +1}$  is formally the same as~\eqref{2.38}, with $\vec S$ having $4 +1$ spin directions of ~\eqref{2.43}, and  the same TCR compatibility kernel of~\eqref{2.33}.

\subsection{Triangle/Oblique (TO) hamiltonian: $N_{OP} =2, N_V =6$}

 The transition is, as in the TCR case,  driven by a  two-component OP \cite{R7,R17,R20}  ${\vec e} \equiv (e_2, e_3)$. Here $N_V = 6$, so we need a square of the cubic term,  $I_3 ^2$ to give six preferred angles. 

{\it i) Scaled free energy, and compatibility potential :}

The scaled Landau  free energy with up to sixth order invariants is~\cite{R17}

\begin{equation}
\displaystyle{\bar{f}_L =(\tau-1)I_2  + I_2 (I_2 - 1)^2  +  C_6( {I_2}^3 - {I_3}^2 )},
\label{2.45}
\end{equation}
where $I_2 = {\vec e}^2\equiv \varepsilon^2$,  $I_3 = e_2^3 - 3e_2 e_3^2$, and $C_6$ is a material constant. 

 In polar coordinates with $\eta_3 = \cos 3 \phi$, this is~\cite{R17}

\begin{equation}
\displaystyle{{\bar f}_L =[ (\tau-1)\varepsilon^2+\varepsilon^2(\varepsilon^2 - 1)^2  ] + C_6 \varepsilon^6 ( 1 - {\eta_3}^2 ).} 
\label{2.46}
\end{equation}

The angular dependence is ${\bar f}_L \sim - \cos 6 \phi$. Minimizing  yields  six martensite variants  with $\sin 6 \phi =0$, at angles    $\phi=  \phi_m \equiv 2 \pi (m-1) / 6$ where $m = 1, 2,.. 6$, where the last term in~\eqref{2.46} vanishes, suppressing the $C_6$ material dependence. The six variants for $\tau < 4/3$
form a hexagon in the $e_2,e_3$ plane, lying on a circle with radius $\varepsilon = \bar \varepsilon (\tau)$  of~\eqref{2.5}. 
%\begin{equation}
%\displaystyle{\bar{ \varepsilon} (\tau)= {\frac{2}{3} ( 1 + \sqrt{1 - 3\tau / 4}) }^{1/2} }.
%\label{2.47}
%\end{equation}

{\it ii) Continuous strains to discrete pseudospins :}

With  the usual  approximation~\eqref{2.35} of $\vec{e} (\vec{r}) \rightarrow \varepsilon (\tau) \vec{S}(\vec{r})$,
the pseudo-spin $\vec S (\vec{r})$ has seven values

\begin{equation}
\vec{S} =  \begin{pmatrix} 0 \\ 0 \end{pmatrix},  \begin{pmatrix} \pm 1 \\ 0 \end{pmatrix},  \begin{pmatrix} \pm \frac{1}{2} \\ \pm \frac{\sqrt{3}}{2}  \end{pmatrix}.
\label{2.47}
\end{equation}

The  Landau term becomes

\begin{equation}
\displaystyle{ {\bar f}_{L} (\tau)  =\bar{ \varepsilon} (\tau) ^2 g_L {\vec S} (\vec r)^2 ~ ; ~ g_L (\tau) \equiv (\tau - 1) + ({\bar \varepsilon}^2-1)^2},
\label{2.48}
\end{equation}
with  $\bar \varepsilon$ of~\eqref{2.5}.

{\it iii) The reduced pseudospin hamiltonian}

The Ginzburg and St Venant terms are as in the TR case. 
The TO case clock-zero  $\mathbb{Z}_{6 +1}$ hamiltonian is as in~\eqref{2.38} with  $\vec S$ having $6 +1$ spin values~\eqref{2.47},  and with the compatibility kernel of ~\eqref{2.33}.

\section{LOCAL MEANFIELD IN TWO SPATIAL DIMENSIONS}

\label{Section3}

With the pseudospin hamiltonians for SR, TCR, SO and TO transitions in hand, we now do local meanfield approximations~\cite{R17} for each of these cases.

\subsection{Square/Rectangle (SR) meanfield : $N_{OP} =1, N_V =2$}

 We write $S(\vec{r}) =  \sigma (\vec{r}) +  \delta S(\vec{r})$,
where $\sigma (\vec{r})  = \langle S(\vec{r}) \rangle$ is the spin  statistical average, and substitute into the hamiltonian~\eqref{2.38}. Retaining only first order terms in $\delta S(\vec{r}) \equiv S (\vec r) - \sigma (\vec r)$, the meanfield hamiltonian is    $H = H_{MF} +  \mathcal{O} (\delta S^2)$. A similar approximation, with identical  final results, can be done in Fourier space, with   ${S}(\vec{k}) =  \sigma (\vec{k}) +  \delta S(\vec{k}) $ substituted in~\eqref{2.39}.

The mean-field hamiltonian is then a sum of a local contribution and a constant,

\begin{equation}
\displaystyle{\beta H_{MF}   \equiv  \sum_{\vec{r}} \beta h_{MF} (\vec r)  - C},
\label{3.2}
\end{equation}  
where

\begin{equation}
\displaystyle{\sum_{\vec{r}}\beta h_{MF} (\vec r ) \equiv \sum_{\vec{r}} V(\vec{r}) S(\vec{r}) =\sum_{\vec{k}} V(\vec{k}) S(\vec{k})^*},
\label{3.2a}
\end{equation}
and $C \equiv \frac{1}{2} \langle \sum  \beta h_{MF} \rangle = \frac{1}{2} \sum_{\vec k} V(\vec k) \sigma(\vec k)^*= \frac{1}{2} \sum_{\vec k} Q_0 |\sigma (\vec k)|^2$.
Here,  $V$ in Fourier and coordinate space is

\begin{equation}
\displaystyle{ \dfrac{V(\vec{k})}{D_0} = [g_L(\tau)  + \xi^2 k^2 +\frac{A_1}{2}  U(\vec{k})]~ \sigma(\vec{k})  }.
\label{3.3a}
\end{equation}  
and
\begin{equation}
\displaystyle{ \dfrac{V(\vec{r})}{D_0} = g_L(\tau) \sigma(\vec{r})  -  \xi^2{\vec  \nabla}^2 \sigma(\vec{r}) +\frac{A_1}{2} \sum_{\vec{r}'} U(\vec{r}-\vec{r}') \sigma(\vec{r}')  }.
\label{3.3b}
\end{equation}  

The meanfield partition function is a product of local contributions

\begin{equation}
\displaystyle{ Z_{MF} =  \sum_{\left\lbrace {S} \right\rbrace } \mathrm{e}^{- \beta H_{MF}} = \prod_{\vec{r}} \sum_{ S (\vec r)} e^{-\beta h_{MF}(\vec{r}) +C}}.
\label{3.4a}
\end{equation}

The self-consistency equation for the statistical average $\sigma(\vec{r})$, with the constant $C$ dropping out, is

\begin{equation}
\displaystyle{ {\sigma}(\vec{r}) = \sum_{ {S} (\vec r) = 0, \pm 1 } {S}(\vec{r}) 
\mathrm{e}^{- \beta V (\vec r) S(\vec r)} / \sum_{ {S} (\vec r)=0, \pm 1 } \mathrm{e}^{- \beta V (\vec r) S(\vec r)}},
\label{3.4b}
\end{equation}
that yields 

\begin{equation}
\displaystyle{ \sigma (\vec{r}) = \dfrac{- 2 \sinh V(\vec{r})}{1 + 2 \cosh V(\vec{r})}  }.
\label{3.5}
\end{equation}

The  equation can also be instructively obtained through the Gibbs-Bogoliubov inequality

\begin{equation}
\displaystyle{ F \leq F_{var} \equiv F_0 + \langle H - H_0 \rangle_0   },
\label{3.6}
\end{equation}

\noindent where the index $0$ refers to an average with a solvable reference system $H_0$,  taken here as $H_0 = - \sum_{\vec{r}} B(\vec{r}) S(\vec{r})$. Here the local field $B(\vec{r})$ is a variational parameter,  and the free energy is  $F_0 = - T \sum_{\vec{r}} \log [ 1 + 2 \cosh\{\beta B(\vec{r})\}]$. The statistical average of $S(\vec{r})$ in the reference system is $\sigma(\vec{r}) \equiv \langle S(\vec{r})\rangle_0 = 2 \sinh \beta B / (1+ 2 \cosh \beta B) $ and the average of $H - H_0$ can also be readily performed since the spins are uncorrelated. The optimal local field $B(\vec{r})$,  that minimizes $F_{var}$ through
$\delta F_{var}/\delta B (\vec{r}) = 0$ is then directly seen as $B(\vec{r}) = - V(\vec{r})$  with the same self-consistency equations as before. Hence $V (\vec{r})$ is indeed the best molecular field to approximate the free energy of the original system.

\begin{figure}[h]
\includegraphics[width=8.0cm]{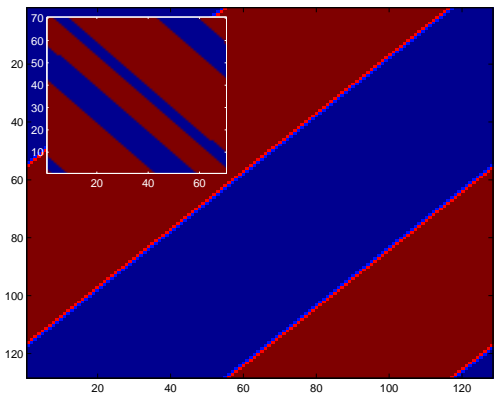}
\caption{Microstructure obtained from the mean-field analysis of the Square-Rectangle (SR) spin model. The parameters are $L=128$, $\xi^2=0.5$, $E_0 =3$, and $\tau = -2.5$, and stiffness $A_1=4$ with  $2 A_1 / A_3 =1$. Twins are oriented along a diagonal as expected. Inset: Twinned ground state from Monte Carlo simulations for the same parameters.} 
\label{twinsMC}
\end{figure}

\paragraph*{}
The mean-field equations have been solved iteratively under a cooling ramp in order to study long-lived glassy states~\cite{R17}. Here we solve  the equations for a fixed constant temperature $\tau$ starting from an initial random configuration.  With an input $\sigma (\vec r)$ and an Fast Fourier Transform (FFT) to a Fourier  $\sigma (\vec k)$, it is easy to find $V(\vec k)$ from the definition~\eqref {3.3a}. A reverse FFT to $V(\vec r)$ is used in~\eqref{3.5} to obtain the next $\sigma (\vec r)$, and the process repeats. Figure~\ref{twinsMC} shows twin microstructure obtained  by solving the mean-field equations, with parameters as in the caption. These twins  are similar to those in experiment~\cite{R8}, to relaxational simulations or to Monte Carlo simulations as in the inset. (Different phases, including certain maze-like textures are also seen in some parameter regimes~\cite{R17}, but do not seem to appear in Monte Carlo simulations.)  

   Thus  a local meanfield approximation  to the pseudospin models is useful to  study  microstructure below  ferroelastic transitions.

 We now  i) study effects of external uniform stress, ii) make contact with the phase diagram of the Blume-Capel model with uniform OP
and iii) show how the Mean-field equations for $\sigma(\vec{r})$ can be obtained through the least-action principle.

\subsubsection{Effects of external stress}

Twins  with oriented, locked-in domain walls of positive energy cost are metastable states, and the uniform single-variant state without domain walls is  the global minimum in free energy. This can be seen by  adding an external stress term $h(\vec r)$  with a simple linear coupling  to the meanfield hamiltonian~\eqref{3.2}:

\begin{equation}
\displaystyle{\beta H_{ext} = - \dfrac{D_0}{2 {\bar \varepsilon} (\tau) } \sum_{\vec r} h ({\vec r})  S ({\vec r} }).
\label{3.7}
\end{equation}

The meanfield self-consistency equations become 

\begin{equation}
\displaystyle{ \sigma (\vec{r}) = \dfrac{- 2 \sinh [ V(\vec{r}) - (D_0/2 {\bar \varepsilon}) h(\vec{r})] }{1 + 2 \cosh [V(\vec{r})  - (D_0/2 {\bar \varepsilon}) h(\vec{r})] }  }.
\label{3.8}
\end{equation}
Starting from random texture seeds with a small uniform external stress, $h = 0.3$, we obtain a uniform state of $ S = \pm 1$ depending on the sign of $h$: the small stress picks out the global minimum. The twins are self-trapped metastable states, that are however  quite rigid against stress:  for a twinned  initial state, a strong stress of about $h = 4$ needs to be applied to destroy the twins and to obtain the uniform ground state. Once the twins have vanished, the system  fails to return to the original  state, i.e.  shows hysteretic behavior.

\subsubsection{Blume-Capel model phase diagram}

 To make contact with treatments of the Blume-Capel model, we suppress the nonlocal couplings by setting $A_1 =0$. The Ginzburg term in~\eqref{2.6} can be recast on a lattice, by setting the gradient to a discrete difference operator $\vec \nabla \rightarrow \vec \Delta$, so that $({\vec \nabla} S)^2 \rightarrow ( {\vec \Delta} S)^2 = 4 S^2 - 2 \sum_{< {\vec r} {\vec r}' >} S (\vec r) S (\vec r ')$. Then the hamiltonian is precisely a Blume- Capel model,  bilinear in the spins (without the  biquadratic term of the Blume-Emery-Griffiths model)~\cite{R22},

\begin{equation}
\displaystyle{H = - J(\tau) \sum_{<{\vec r} {\vec r} ' >}S (\vec r)  S (\vec r ') + \Delta (\tau) \sum_{\vec r} S (\vec r) ^2},
\label{3.9}
\end{equation}

\begin{figure}[h]
\includegraphics[width=8.0cm]{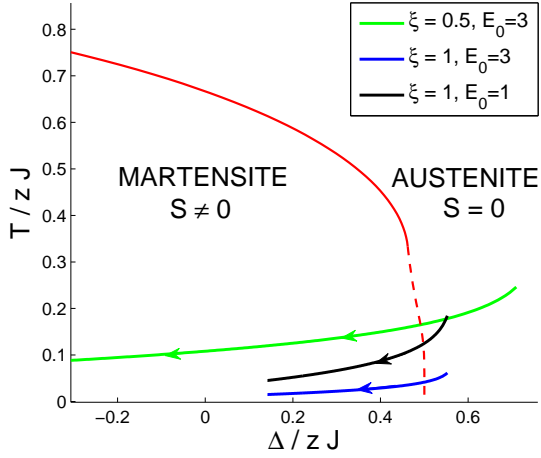}
\caption{Mean-field phase diagram for the spin-1 model for the Square-Rectangle (SR) transition. The red solid line represents second-order phase transition and the red dashed line represents a first-order phase transition \cite{R22}, meeting at a tricritical point. The crystal field $\Delta (\tau)$ depends on the temperature $\tau$, so cooling is a  phase-diagram trajectory. Three directed trajectories with different sets of parameters are shown, for cooling from $\tau=4/3$ to $\tau=-2.5$.The lines intersect the first-order  transition line (dashed) for $\tau \simeq 1$.}
\label{PDiagram}
\end{figure}

\noindent   There is  temperature-dependence in the on-site crystal field term $\Delta  (\tau) \equiv  D_0 (\tau) [ g_L  (\tau) + 4 \xi^2 ]/(2 \beta)$, and in  the ferromagnetic  coupling             $J (\tau) \equiv D_0 (\tau)  \xi^2/ \beta$.

\begin{figure}[h]
\includegraphics[width=8.0cm]{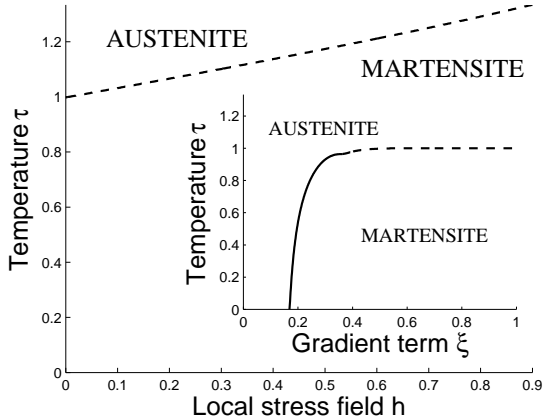}
\caption{ Mean-field phase diagram of the Square-Rectangle (SR) transition in the ($h$,$\tau$) plane for $\xi =1$. The $h=0$ first-order phase transition temperature of $\tau=1$ moves up with applied stress $h$. Inset: Phase diagram in the ($\xi$,$\tau$) plane.  For $\xi$ large , there is a first-order phase transition line (dashed) with  $\tau =1$, while  for $\xi$ small,  there is a second-order  transition line (solid), that moves to lower temperature.} 
\label{8}
\end{figure}

The model  can be  studied within the (uniform)  mean-field approximation. An expansion of the mean-field free energy yields an analytical expression for the line of critical points    $T_c = z J/3$; and the location of the tricritical point,  $\Delta_c = \frac{2}{3} z J \log 2$, where $z=4$ is the number of nearest neighbors. Figure~\ref{PDiagram} shows the well-known phase diagram of this model. Both $\Delta$ and $J$ depend on the temperature $\tau$, so although a given temperature corresponds to a point, a cooling path is a line in the phase diagram. These lines intersect the first-order transition curve for $\tau \simeq 1$, the Landau transition temperature between  ' paramagnetic ' austenite and ' ferromagnetic ' martensite. Figure~\ref{8} shows the meanfield phase diagram for two parameter planes $(h, \tau)$ and $(\xi, \tau)$. In a certain range of parameters, the spin model is consistent with the Landau theory that predicts a first-order phase transition at $\tau = 1$.

\subsubsection{Field theory for $\sigma (\vec{r})$}

 We show here how the partition function may be transformed to obtain a field theory for  $\sigma(\vec{r})  = \langle S(\vec{r}) \rangle$, so the mean-field equation~\eqref{3.5} results from a saddle-point approximation of a functional integral. Other mean-field equations in this paper can similarly be obtained as saddle point approximations of field theories.

The partition function can be compactly written as
\begin{equation}
\displaystyle{Z = \sum_{\left\{S \right\}} \exp \left( \frac{1}{2} \sum_{\vec{r} \vec{r}'} S_{\vec{r}} K_{\vec{r} \vec{r}'} S_{\vec{r}'} \right) },
\end{equation}
where \mbox{$K_{\vec{r} \vec{r}'} = \beta J(\tau) \delta_{< \vec{r},\vec{r}'>} - \frac{D_0 A_1}{2} U(\vec{r}-\vec{r}') - 2 \beta \Delta(\tau) \delta_{\vec{r},\vec{r}'}$}. The first Kronecker symbol is non-zero only if $\vec{r}$ and $\vec{r}'$ are neighbors. We note that the kernel 
$K_{\vec{r} \vec{r}'}$ can be recast using usual matrix notations
\begin{equation}
\displaystyle{K_{\vec{r} \vec{r}'} = D_0 \Braket{\vec{r}| \xi^2 \nabla^2 - \frac{A_1}{2} U - g_L (\tau) | \vec{r}'}},
\end{equation}
where $ \langle \vec{r}| U | \vec{r}' \rangle = U(\vec{r}-\vec{r}')$. We may then use the standard Hubbard-Stratonovich transformation $\int e^{- \sum _{ij} A_{ij} x_i x_j+ \sum_i B_i x_i} d^nx=\sqrt{ \frac{\pi^n}{\det{A}} }e^{\frac{1}{4}B^TA^{-1}B}$ to find the exact integral representation of the partition function
\begin{equation}
\displaystyle{Z = \frac{1}{(2 \pi)^{N/2} \sqrt{\det K}} \int \prod_{\vec{r}} d \phi(\vec{r}) \mathrm{e}^{- S [ \phi ]} },
\end{equation}
with the action
\begin{equation}
\displaystyle{ S[ \phi] = \frac{1}{2} \sum_{\vec{r} \vec{r}'} \phi(\vec{r}) K_{\vec{r} \vec{r}'}^{-1} \phi(\vec{r}') - \sum_{\vec{r}} \mathrm{log} \left( 1 + 2 \cosh(\phi(\vec{r})) \right) }.
\end{equation}
Finally, we define $\sigma (\vec{r}) = \sum_{\vec{r}'} K^{-1}_{\vec{r} \vec{r}'} \phi (\vec{r}')$. The partition function reads
\begin{equation}
\displaystyle{Z =  \int \mathcal{D} [\sigma(\vec{r})]  \mathrm{e}^{- S [ \sigma ]} },
\end{equation}
where we have defined the formal measure $\mathcal{D} [\sigma(\vec{r})] = \sqrt{\det{K}} / (2 \pi)^{N/2} \prod_{\vec{r}} d \sigma(\vec{r})$.
With this  field-theoretical formulation of the partition function, our problem, the Mean-field approximation is obtained by  minimizing the action 
\begin{align}
\displaystyle S[ \sigma] &= \frac{1}{2} \sum_{\vec{r} \vec{r}'} \sigma(\vec{r}) K_{\vec{r} \vec{r}'} \sigma(\vec{r}') \\ \notag
\displaystyle &- \sum_{\vec{r}} \mathrm{log} \left( 1 + 2 \cosh(\sum_{\vec{r}'} K_{\vec{r} \vec{r}'} \sigma(\vec{r}')) \right) .
\end{align}

Within the saddle-point approximation, $\sigma(\vec{r})  = \langle S(\vec{r}) \rangle$ so that the field $\sigma$ is indeed the statistical average of the spin. We note that $V(\vec{r}) = -\sum_{\vec{r}'} K_{\vec{r} \vec{r}'} \sigma(\vec{r}')$ (eq.~\eqref{3.3b}), so the minimization of the action yields  the mean-field equation~\eqref{3.5}, as expected.

\subsection{Triangle/Centred-Rectangle (TCR) meanfield:  $N_{OP} =1, N_V =3$}

The TCR case spin hamiltonian is~\eqref{2.38}, with spin values $\vec S = (0,0), (1,0) (-1/2, \pm \sqrt{3} /2)$ of~\eqref{2.36} and $g_L$ as in~\eqref{2.37}. 
Since $N_{OP} =2$ for the TCR, SO, TO and CT transitions, their meanfield equations are all formally the same.
 From the substitution $\vec{S_\ell }(\vec{r}) =  \vec{\sigma_\ell }(\vec{r}) +  \delta \vec{S}(\vec{r})$ and linearization in                                                                                                                           $\delta \vec{S}(\vec{r}) \equiv \vec{S_\ell }(\vec{r})-  \vec{\sigma_\ell }(\vec{r})$  the meanfield hamiltonian is $\beta H_{MF} \equiv \sum_{\vec r} \beta h_{MF} (\vec r) -C$, as in (49) but  the local contribution is now

 \begin{equation}
\displaystyle{  \sum_{\vec{r}} \beta h_{MF} (\vec r)  =  \sum_{{\vec r} , \ell =2, 3} V_{\ell} (\vec{r}) S_{\ell}(\vec{r}) = \sum_{{\vec k} ,\ell} V_{\ell} (\vec{k}) S_{\ell}(\vec{k})^{*} }
\label{3.10}
\end{equation} 
  and $C \equiv  \frac{1}{2} \sum \langle \beta h_{MF} \rangle = \frac{1}{2} \sum_{\vec k , \ell} V_\ell (\vec k) \sigma_\ell (\vec k)^*$.  
The functions $V_2$ and $V_3$ are defined in Fourier space by

\begin{equation}
\displaystyle{ V_2 (\vec{k}) = P_2^0 (\vec{k}) \sigma_2(\vec{k}) +\frac{D_0 A_1}{2} U_{2 3}(\vec{k})\sigma_3(\vec{k})  },
\end{equation}

\begin{equation}
\displaystyle{ V_3 (\vec{k}) = P_3^0 (\vec{k}) \sigma_3 (\vec{k}) +\frac{D_0 A_1}{2} U_{32}(\vec{k})\sigma_2(\vec{k})  }.
\end{equation}
where 
\begin{equation}
\displaystyle{ P_\ell ^0 (\vec{k}) = D_0 [ g_L + \xi^2 \vec{k}^2 + \frac{A_1}{2} U_{\ell \ell}(\vec{k}) ]  }.
\label{3.11}
\end{equation}
 
Defining  $P_\ell$ in Fourier space as $ P_\ell (\vec{k}) \equiv  P_\ell ^0 (\vec{k})  \sigma_\ell (\vec{k})$, the  coordinate space meanfield hamiltonian of~\eqref{3.10} is then 
  
\begin{align}
\displaystyle \beta h_{MF} (\vec r) & =  \sum_{\ell} P_{\ell} (\vec{r}) S_{\ell}(\vec{r}) \\ \notag
\displaystyle &+ \dfrac{D_0 A_1}{2} \sum_{\vec{r}'} U_{2 3}(\vec{r}-\vec{r}') [\sigma_{2}(\vec{r}) S_{3}(\vec{r}') + \sigma_{3}(\vec{r}') S_{2}(\vec{r}) )].
\label{3.10}
\end{align} 
The partition function of this linearized meanfield  Hamiltonian  can again be factorized as in~\eqref{3.4a}.

%\begin{equation}
%\displaystyle{ Z_{MF} =  \sum_{\left\lbrace \vec{S} \right\rbrace } \mathrm{e}^{- \beta H_{MF}} = \prod_{\vec{r}} \sum_{\vec S (\vec r)} e^{-\beta h_{MF}(\vec{r}) + C}  }.
%\end{equation}

The self-consistency equations  as in~\eqref{3.4a},~\eqref{3.4b} for the statistical averages $\left\lbrace \vec{\sigma}(\vec{r})\right\rbrace $ again have the constant $C$ cancelling, so now with              $\vec V = (V_2 , V_3 )$ and $\vec S = (S_2 , S_3)$, 

\begin{equation}
\displaystyle{ \vec{\sigma}(\vec{r}) = \sum_{ \vec{S} (\vec r) } ~\vec{S}(\vec{r}) 
\mathrm{e}^{- {\vec V} (\vec r) . {\vec S} (\vec r)} / \sum_{ \vec{S} (\vec r) } \mathrm{e}^{- {\vec V} (\vec r) . {\vec S} (\vec r)}} .
\label{3.12}
\end{equation}

In terms of the $N_V$ variant states $\vec S = (\cos \phi_m ,  \sin \phi_m )$ with $m = 1,2 ...N_V$ this can be formally expressed for TCR, SO,  TO and CT cases as 

\begin{equation}
\displaystyle{ \sigma_2 = \dfrac{\sum_{m=1}^{N_V} \cos \phi_m \mathrm{e}^{-(\cos \phi_m V_2 + \sin \phi_m V_3)}}{1 + \sum_{m=1}^{N_V} \mathrm{e}^{-(\cos \phi_m V_2 + \sin \phi_m V_3)}}},
\label{3.12a}
\end{equation}

\begin{equation}
\displaystyle{ \sigma_3= \dfrac{\sum_{m=1}^{N_V} \sin \phi_m \mathrm{e}^{-(\cos \phi_m V_2 + \sin \phi_m V_3)}}{1 + \sum_{m=1}^{N_V} \mathrm{e}^{-(\cos \phi_m V_2 + \sin \phi_m V_3)}}}.
\label{3.12b}
\end{equation}

For the TCR case sums over the $N_V =3$  spin values of~\eqref{2.36}, this is

\begin{equation}
\displaystyle{ \sigma_2= \dfrac{1}{2} \dfrac{\textrm{e}^{-\frac{3}{2}V_2} - \cosh(\frac{\sqrt{3}}{2} V_3)}
 {[\textrm{e}^{-V_2 } \cosh(\frac{V_2 }{2} ) + \cosh(\frac{\sqrt{3}}{2} V_3)]}},
\label{3.13a}
\end{equation}

\begin{equation}
\displaystyle{ \sigma_3= - \dfrac{\sqrt{3}}{2} \dfrac{\sinh(\frac{\sqrt{3}}{2} V_3)}
 {[\textrm{e}^{-V_2} \cosh(\frac{V_2}{2} ) + \cosh(\frac{\sqrt{3}}{2} V_3)]}}.
\label{3.13b}
\end{equation}
where the position dependences of $\sigma_\ell (\vec r)$ and $V_\ell (\vec r)$ are left implicit.
The coupled equations~\eqref{3.13a},~\eqref{3.13b} were solved iteratively on a $ L \times L =256 \times 256$ lattice with periodic boundary conditions with parameter values $\xi^2 = 0.8$, $A_1 = 5$,       $\tau = - 6.5$, and  $E_0 = 0.01$. Here, and throughout the following other cases,  $T_0 = 1.0$, $T_c = 0.9$.

\begin{figure}[h]
\includegraphics[width=8.0cm]{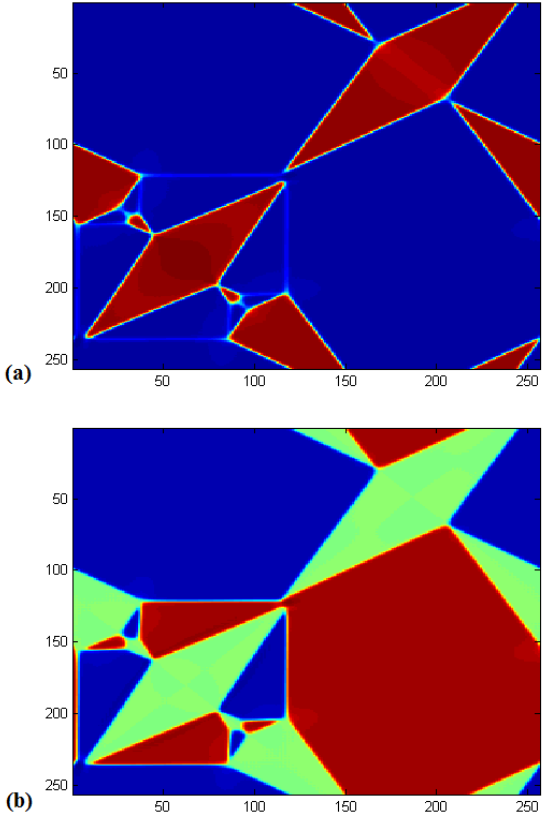}
\caption{Final-state microstructure obtained from meanfield self-consistency equations for the Triangle-Centred Rectangle (TCR) transition:  (a)  strain 
component $e_2$ ;  (b) strain component $e_3$. The color code of red (blue) corresponds to positive (negative) values,  and
green to zero. Note the sharp domain walls.  Parameters are $L=256$, $\xi^2 = 0.8$, $E_0 = 0.01$,  scaled temperature $\tau = -6.5$ and stiffness $A_1 = 5$.} 
\label{stars}
\end{figure}

Figure~\ref{stars} shows the relaxed microstructure obtained after $10^5$ iteration steps. As in continuous-variable simulations in strains or displacements~\cite{R4,R7} we also obtain nested-star patterns as observed in experiments~\cite{R8} for lead orthovanadate. However, unlike the continuous-variable models which are computationally intensive, the spin models and the local meanfield solutions reach the complex microstructure relatively rapidly.

\subsection{ Square/Oblique (SO) meanfield: $N_{OP} =2, N_V =4$}

The SO case spin Hamiltonian  is formally the same as~\eqref{2.38} but with SO case spin  values $\vec{S} = (\pm 1/ \sqrt{2},\pm 1/{2})$ of~\eqref{2.43}, and $g_L$ is as in~\eqref{2.44}.  Doing a local meanfield approximation as before, the formal self-consistency equations of ~\eqref{3.12a}, ~\eqref{3.12b} become

\begin{equation}
\displaystyle{ \sigma_2 = -\dfrac{1}{\sqrt 2} \dfrac{2 \sinh(\frac{V_2+V_3}{\sqrt{2}})+2 \sinh(\frac{V_2 -V_3}{\sqrt{2}})}
 {[1+ 2 \cosh(\frac{V_2 +V_3 }{\sqrt{2}})+ 2\cosh(\frac{V_2 -V_3}{\sqrt{2}})]}},
\end{equation}

\begin{equation}
\displaystyle{ \sigma_3 = -\dfrac{1}{\sqrt 2} \dfrac{2 \sinh(\frac{V_2 +V_3 }{\sqrt{2}})-2 \sinh(\frac{V_2 -V_3 }{\sqrt{2}})}
 {[1+  2 \cosh(\frac{V_2 +V_3}{\sqrt{2}})+2 \cosh(\frac{V_2 -V_3}{\sqrt{2}})]}}.
\label{e_MFSO2}
\end{equation}

\begin{figure}[h]
\includegraphics[width=8.0cm]{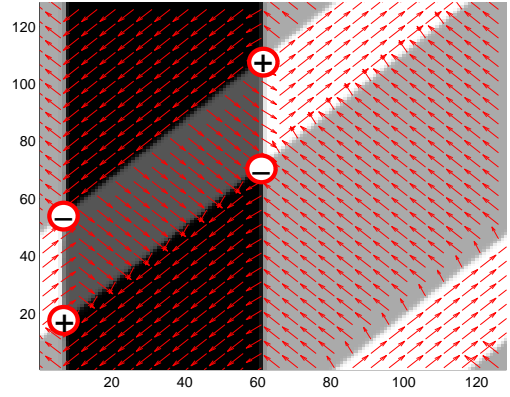}
\caption{Final state microstructure from meanfield equations  for the Square-Oblique (SO).  The four variants are in four different colors, and pseudospin spins orientations are also denoted. The discretized vortices (topological charge $+1$) and anti-vortices (topological charge $-1$), expected in the classical clock models, are also identified. 
Parameters used are $L=128$, $\xi^2 = 0.3$,  $E_0 = 0.2$,  and $\tau = -2.5$  and  $A_1 = 6$. } 
\label{SO_microstructure}
\end{figure}

The coupled equations were solved iteratively on a $ 128 \times 128 $ lattice with periodic boundary conditions and  for different temperatures $\tau$, starting from an initial random texture. Figure~\ref{SO_microstructure} shows  that the microstructure obtained for $\tau = -2.5$, has  {\it  vortices}, as in the classical clock models or in the XY model. This  vortex in the strain  field differs of course,  from an edge dislocation that is a structural defect in the displacement field. The pseudospin vortex at the meeting point of domain walls is characterized by the  winding number or topological charge

\begin{equation}
\displaystyle{q_i = \frac{1}{2 \pi} \oint_{\Gamma_i} \vec{\nabla} \theta . d \vec{r}},
\end{equation}
where $\theta (\vec{r})$ is the polar angle of the spin $\vec{S}(\vec{r})$, that equals $\phi_m$ in the variant regions,  and $\Gamma_i$ is an arbitrary contour surrounding the $i$-th vortex. The topological charge is $q_i=1$ for a vortex and $q_i =-1$ for an anti-vortex. Thanks to the periodic boundary conditions, we have $\sum_i q_i = 0$.  Vortex solutions for complex fields at three-domain meeting points have been considered~\cite{R23}.

\begin{figure}
\includegraphics[width=8.0cm]{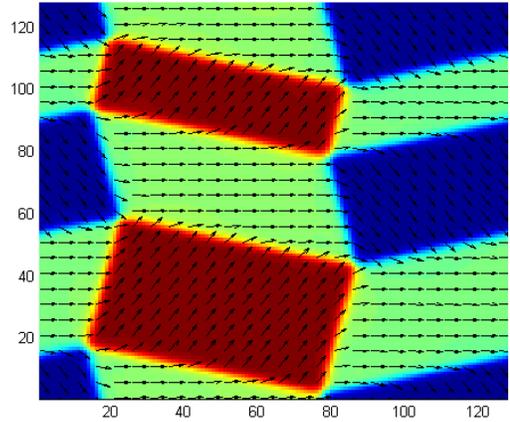}
\caption{Final state microstructure  from meanfield equations for the Triangle-Oblique (TO). The three variants are
in three different colors, and spin orientations are also shown. Only three of the six variants (see Fig 8),  finally survive. Parameters used are $L=128$, $\xi^2 = 0.35$, $E_0 = 0.2$, scaled temperature  $\tau = -2.6$, and  stiffness $A_1 = 5$.} 
\label{TOfig}
\end{figure}

\subsection{Triangle/Oblique (TO) meanfield: $N_{OP} =2, N_V = 6$}

\begin{figure}[h]
\includegraphics[width=8.0cm]{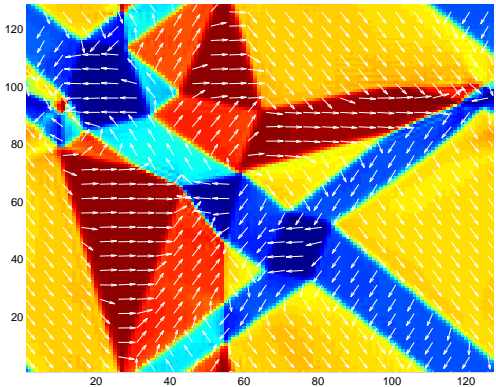}
\caption{Transient state for the (TO) transition  with the six variants with different colors, with parameters as for Fig 7.} 
\label{TOfigtransient}
\end{figure}

The  TO case hamiltonian is as in~\eqref{2.38} but with TO case spin values  $\vec S = (0,0), (\pm 1, 0), (\pm 1/2, \pm \sqrt{3} /2)$ from ~\eqref{2.47}, and $g_L$ is as in~\eqref{2.48}.
The general meanfield self-consistency equations of~\eqref{3.12a}, ~\eqref{3.12b} are then

\begin{equation}
\label{}
\displaystyle{ \sigma_2 = - \dfrac{ 2 \sinh(V_2)+\sinh(I)+\sinh(J)}
 {1+ 2 \cosh(V_2)+2 \cosh(I)+ 2\cosh(J)}},
\end{equation}

\begin{equation}
\displaystyle{ \sigma_3 = -\frac{\sqrt{3}}{2} \dfrac{2 \sinh(I) - 2 \sinh(J)}
 {1+ 2 \cosh(V_2)+2 \cosh(I)+ 2 \cosh(J)}},
\label{}
\end{equation}
where $I = (V_2 + \sqrt{3} V_3)/2$ and $J = (V_2 - \sqrt{3} V_3)/2$.  
Figure~\ref{TOfig} shows the ground state obtained from these coupled meanfield equations with parameters $L=128$, $\xi^2 = 0.35$, $E_0 = 0.2$, $\tau = -2.6$ and $A_1 =5$.  
We note that discrete vortices at the junction of the six martensite variants, are  seen only during the iterations through transient states as in  Figure~\ref{TOfigtransient}. The final state microstructure  shows no vortices, and only three out of the six variants finally remain,  bounded by nonintersecting domain walls, as the other variants  vanish during the course of the textural evolution.  The suppression of   vortices  at least for these parameter values,  could be due to the energy costs of the gradient and powerlaw terms.

\section{PSEUDOSPIN HAMILTONIAN AND LOCAL MEANFIELD IN THREE SPATIAL DIMENSIONS}

\label{Section4}

We outline the hamiltonian derivations for the cubic/tetragonal case and then do a meanfield analysis. The approach can also be followed for other 3D transitions~\cite{R17}.

\subsection{Cubic/Tetragonal  (CT) hamiltonian: $N_{OP} = 2, N_V =3$}

For the cubic-to-tetragonal  or 'CT' transition, the symmetry-adapted strains are the dilatation $e_1=(1/ \sqrt{3}) (e_{xx}+ e_{yy}+ e_{zz})$, the two deviatoric OP strains $e_2=(1/ \sqrt{2}) (e_{xx}-e_{yy})$, $e_3=(1/ \sqrt{6}) (e_{xx}+e_{yy}-2 e_{zz})$, and the three shear strains $e_4 = 2 e_{yz}$, $e_5 = 2 e_{zx}$, $e_6 = 2 e_{xy}$.

The OP components are  the two deviatoric strains $\vec e = (e_3 , e_2)$, and the remaining four non-OP compressional and shear  strains are  $e_1, e_4, e_5, e_6$.
The Landau free energy invariant under symmetries of the cubic unit-cell, was originally given by
Barsch and Krumhansl~\cite{R10}, where the cubic invariant is now $ I_3 =  (e_3^3-3 e_3 e_2^2)$, and in scaled form is 

\begin{equation}
\displaystyle f_{L}  =   \tau (e_2^2+e_3^2) - 2 (e_3^3-3 e_3 e_2^2) + (e_2^2+e_3^2)^2 .
\end{equation}
The Ginzburg term is formally identical to~\eqref{2.29} but in 3D.   

The non-OP terms, harmonic in the four remaining physical strains are 
 
\begin{equation}
\displaystyle{  f_{non } = \frac{A_1}{2}  e_1^2 + \frac{A_4}{2} ( e_4^2 + e_5^2 + e_6^2)}.
\end{equation}
and are minimized subject to the compatibility constraint~\eqref{2.8} in 3D. There are six equations, from cyclic permutations of the labels $x,y,z$ of the two equations

\begin{equation}
\displaystyle{ 2\partial_{x} \partial_y e_{yz} -  \partial^2_{z} e_{yy} - \partial^2_{y} e_{zz} = 0},
\end{equation}
\begin{equation}
\displaystyle{  \partial_{y} \partial_z e_{x x} +  \partial^2_{x} e_{y z} - \partial_{x} \partial_y e_{z x} - \partial_{x} \partial_z e_{x y} = 0}.
\end{equation}
By going to Fourier space one finds the second set is an identity,  if the first set is satisfied. These constraint equations can be recast in terms of the symmetry-adapted strains $e_1, e_2...e_6$.  Minimizing ${\bar F}_{non}$ with these constraints (either through Lagrange multipliers~\cite{R7}, or through direct solution for  $e_4, e_5, e_6$  and minimization~\cite{R17} in the remaining $e_1$),  yields the non-OP strains  in terms of the  OP strains $e_2$ and $e_3$. Substitution into the harmonic non-OP free energy yields the compatibility term 

\begin{equation}
\displaystyle{ {\bar F}_{compat}  = \frac{1}{2} \sum_{\ell,\ell' = 2, 3} \sum_{\vec{k}} A_1 U_{\ell \ell'}(\vec{k})e_\ell (\vec{k})  e_{\ell '}(\vec{k})^{*}},
\label{Kernels3D}
\end{equation}
where the kernels $U_{\ell,\ell'} (\vec k)$ in Fourier space~\cite{R17} are  given in the Appendix.

The procedure is formally just as in the TCR case, as  the CT case also has the same   $N_{OP} = 2, N_V =3$,  pseudo-spin values $\vec{S} = (0,0), (1,0), (-1/2, \pm \sqrt{3}/2)$ and again $\bar \varepsilon (\tau)=  3/4 ( 1 + \sqrt{1 - 8 \tau / 9})$. The spatial dimension
only enters in the 3D compatibility potential of kernels~\eqref{Kernels3D}, and in the 3D lattice positions  $\vec r = (x, y, z)$ and Brillouin zone wave-vectors $\vec k = (k_x, k_y, k_z)$.

\subsection{ Cubic/Tetragonal meanfield : $N_{OP} =2, N_V =3$}

\begin{figure*}[tbph] 
\begin{center}
\includegraphics[width=16.0cm]{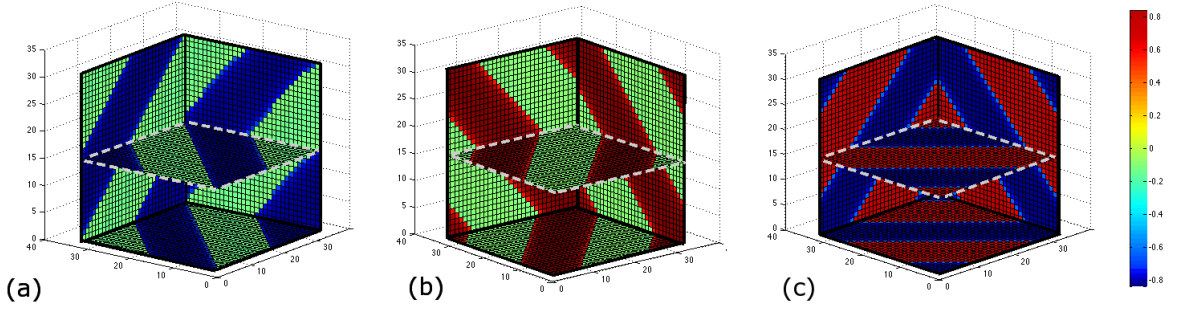}
\caption{Twins in the (111) plane obtained from the meanfield self-consistency equations for the 3D cubic-to-tetragonal transformation. The color bar represents $S_y$. The parameters are $L=32$, $\xi^2=10$, $E_0 = 0.001$, scaled temperature $\tau = -0.5$  and stiffnesses $A_1 = 4.8$, $A_4 = 2.4$. The microstructures (a),(b), and (c) show three different twin orientations obtained, for different runs.} 
\label{twins}
\end{center}
\end{figure*}

We numerically solved the CT meanfield equations~\eqref{3.13a},~\eqref{3.13b} that are  as for the TCR  case,  with the  kernels as in the Appendix.  We took  a $32 \times 32 \times 32$ lattice with periodic boundary conditions, and parameters  $\xi^2=10$, $E_0 = 0.001$,  $T_0 = 1, T_c = 0.9$, and stiffnesses $A_1 = 4.8$, $A_4 = 2.4$.  Fourier transforms enable a computation at each step, of  the functions $V_2(\vec{r})$ and $V_3 (\vec{r})$. Figure~\ref{twins} shows the microstructure, with twins at diagonal orientations, as found in continuous-variable simulations.

\section{OTHER RELATED MODELS}

\label{Section5}

Modified truncations of these structural-transition free energies can induce other hamiltonians, that can be studied  purely as interesting spin models in statistical mechanics.

Let us suppress the zero state, and fix only the circle radius $\varepsilon \rightarrow {\bar \varepsilon} (\tau)$,  while keeping all continuous polar angles, now denoted by $\theta (\vec r)$, with  values  $2 \pi > \theta  \geq 0$,

\begin{equation}
\vec{e} (\vec{r})=  \begin{pmatrix} e_2(\vec{r}) \\ e_3(\vec{r}) \end{pmatrix}  \rightarrow {\bar  \varepsilon} (\tau) \begin{pmatrix} \cos \theta (\vec{r}) \\ \sin \theta (\vec{r}) \end{pmatrix}
\end{equation} 
so  as in the XY model of planar spins, ${\vec S} = (\cos \theta, \sin \theta)$. The Ginzburg discrete-difference term of~\eqref{2.6} then induces an XY- like  ferromagnetic interaction. The Landau free energy in polar coordinates in all cases has angular dependence ${\bar f}_L \sim - B \cos N_V \phi$ as in ~\eqref{2.27}, ~\eqref{2.42}, ~\eqref{2.46}. Putting all this together, 
the free energy induces an XY ferromagnet model with a  long-range potential, and a symmetry-breaking local field :  

\begin{align}
\displaystyle \beta H & = \frac{D_0}{2} [-  B \sum_{\vec r} \cos [N_V \theta (\vec r)]  \notag \\
                      & -2 \xi^2 \sum_{<\vec{r},\vec{r}'>} \cos(\theta(\vec r) - \theta(\vec r ')) ]+ \beta H_C (\{ \theta\})
\end{align}

 Here as ${\vec S}^2  =1$, there is no quadratic local term, and  $\beta H_C (\{ \theta \})$ is a term coupling  the continuous-angle variables                                                                                     $\cos \theta (\vec r)$ and $\sin \theta  (\vec r)$, 

 \begin{align}
 \displaystyle \beta H_C  [\{\theta \} ] &  = \frac{A_1 D_0}{4} \sum_{{\vec r}, {\vec r}'}  U_{22} \cos \theta  (\vec r) \cos \theta  (\vec r ') \notag \\
 &+ U_{33} \sin \theta (\vec r) \sin \theta(\vec r ') \notag \\ 
\displaystyle & + U_{23} \cos \theta (\vec r) \sin \theta (\vec r ')  + U_{32} \sin \theta (\vec r) \cos \theta  (\vec r ') 
 \end{align}
  
 %Even for $A_1 = 0$, the $O(2)$ symmetry of the usual XY Hamiltonian is here broken into the discrete group of symmetry $\mathbb{Z}_{3} = \mathbb{Z} / 3 \mathbb{Z}$ because of the external potential. For a system of $N$ spins, the 
 \noindent and a partition function

\begin{equation}
\displaystyle{ Z = \int_{[0, 2\pi]^N} \prod_{\vec{r}} d \theta (\vec{r}) \exp(- \beta H[\{\theta (\vec{r})\}; \tau])}.
\end{equation} 
This model includes all angles, even away from minima, and so can describe  slowly transiting states across $N_V$ saddle points, as  in experiment~\cite{R24}.

 Similar XY models with symmetry-breaking fields  (without the powerlaw interaction) have been studied~\cite{R24}. A dual transform in that case extracts the topological vortices with logarithmic interactions, and in this model could also induce a powerlaw anisotropic vortex interaction. A real-space renormalization group analysis of the 2D Coulomb gas as in the Kosterlitz -Thouless transition is well-known~\cite{R25}, and  could be repeated for this model.  Renormalization flows in the context of martensitic transitions have been studied in other models~\cite{R26}.

For strong symmetry-breaking in minima angular directions  ( $|B|$ large), the continuous angle $\theta$ become discrete and takes on values that we denote as $\theta \rightarrow  \phi =\phi_m$, and one gets pure clock models ($\mathbb{Z}_{N_V}$)  with a Hamiltonian that now has powerlaw potentials,

\begin{equation}
\displaystyle{\beta H_{clock} = - D_0 \xi^2 \sum_{<{\vec r} {\vec r'}>} \cos[\phi (\vec r) - \phi (\vec r ')] +  \beta H_C (\{\phi \})}.
\end{equation} 
We can even make one more approximation by reducing the XY interaction to a Kronecker-delta coupling, yielding a $q$-state  Potts~\cite{R19} model with $q = N_V$:  

\begin{equation}
\displaystyle{\beta H_{Potts} = - D_0 \xi^2 \sum_{<{\vec r} {\vec r '}>} \delta_{\vec{S} (\vec r) ,\vec{S} (\vec r ')} +\beta H_C (\{\phi \}) }.
\end{equation} 
Potts hamiltonians with large number of spin components  $q$ have been studied as models for configurational glasses \cite{R19}.

\section{Conclusion}

\label{Section6}

A standard approach to obtaining  microstructure  of structural transitions is to solve 
evolution equations  for relaxation to a minimum,  in continuous variables such as displacements, phase fields or strains~\cite{R4,R5,R6,R7}.  We have here considered 
the reduced hamiltonian models in {\it discrete} pseudospins describing four structural transitions in two dimensions, as well as the three dimensional cubic-to-tetragonal transition. These 'clock-zero' models  have a zero state as well as clock states, and the pseudospin hamiltonian has  an on-site term, an exchange interaction and a powerlaw interaction term.  For the square/rectangle  case, the pseudo-spin model without powerlaw interactions corresponds to the Blume-Capel  spin-1 model with temperature-dependent couplings. Using a local meanfield approach, we have obtained the microstructure  for 2D and 3D transitions, as obtained in continuous-variable strain dynamics. For example, the characteristic nested star microstructure of the triangle  transition emerges easily from the meanfield solution. 

The textures  of the square/oblique  (SO)  and triangle/oblique  (TO)  transitions, with $N_V =4, 6$, which have not been previously studied, include  vortex configurations of the $\mathbb{Z}_{N_V + 1}$  clock models, at intersections between variant domain walls.  The SO final microstructure has positive/negative vortices in regular patterns, and all four variants are present. For the  TO case, at least for particular parameters, we find the six-variant-vortices appear only as transient solutions, with the final state  having  no vortices, with only non-intersecting closed-domains of  three variants.  Finally,  for the three dimensional  cubic/tetragonal transition, we obtain the diagonal twinning that is consistent with  previous studies~\cite {R4,R6,R7}.  In all cases, the local meanfield final microstructure emerges relatively rapidly, compared to the  slow evolution towards steady-state of the continuum differential-equation dynamics.

Further work can involve studies of  pseudospin hamiltonians~\cite{R17} for other structural transitions in 2D and 3D  in the local meanfield approach. By including quenched disorder, such pseudo-spin models may be used to study strain glass behavior in martensitic alloys~\cite{R21},  and relate solutions to the tweed precursors~\cite{R5} in analogy with spin-glass like behavior, and to random-field models~\cite{R16}. Monte Carlo simulations can be used to study martensitic nucleation and growth~\cite{R27}. Other related spin models  of interest in their own right may include geometric nonlinearities that
yield complex heirarchical-twin patterns~\cite{R2,R4}. 
%Extensions of the strain-representation scaled free energies to higher order in the spontaneous strain, can systematically include geometric nonlinearities,  as a further  topic of investigation.

In conclusion, the discrete-variable pseudospin model in local meanfield approximation, is therefore a useful approach to the study of martensitic texturing.
            
\begin{acknowledgments}
We are grateful to the Center for Nonlinear Science at Los Alamos National Laboratory for the award of
a summer studentship in 2009 to RV. We acknowledge useful discussions with Marcel Porta and Avadh Saxena. This work was carried out under the auspices of the National Nuclear Security Administration of the U.S. Department of Energy at Los Alamos National Laboratory under Contract No. DE-AC52-06NA25396. NSERC of Canada and ICTP, Trieste,  is also thanked for support.
\end{acknowledgments}

\appendix

\section{Kernels for the cubic-to-tetragonal transition}

In this Appendix we state the explicit form of the bulk kernels $U_{\ell \ell'}$ obtained elsewhere~\cite{R17}  for the 3D cubic-to-tetragonal transition. To do so we define the coefficients $O^{(s)}_{\alpha}$ and $O_{\alpha}$  by

\begin{equation}
\displaystyle{O_1^{(4)} = \frac{-1}{\sqrt{3}} (k_y^2+k_z^2), O_2^{(4)} =  \frac{k_z^2}{\sqrt{2}}, O_3^{(4)} =  \frac{1}{\sqrt{6}} (2 k_y^2 - k_z^2)},
\end{equation} 
\begin{equation}
\displaystyle{O_1^{(5)} = \frac{-1}{\sqrt{3}} (k_x^2+k_z^2), O_2^{(5)} =  \frac{-k_z^2}{\sqrt{2}}, O_3^{(5)} =  \frac{1}{\sqrt{6}} (2 k_x^2 - k_z^2)},
\end{equation} 
\begin{equation}
\displaystyle{O_1^{(6)} = \frac{-1}{\sqrt{3}} (k_y^2+k_x^2), O_2^{(6)} =  \frac{1}{\sqrt{2}} (k_x^2-k_y^2), O_3^{(6)} =  \frac{-1}{\sqrt{6}} (k_x^2+k_y^2)},
\end{equation} 

\begin{equation}
\displaystyle{O_4 = k_y k_z , O_5 = k_x k_z, O_6 = k_x k_y},
\end{equation} 

Let $ \bar{O}^{(s)}_{\alpha} = O^{(s)}_{\alpha} / O_{s}$ and $G_{\alpha \beta} = \sum_s (A_s/A_1) \bar{O}^{(s)}_{\alpha} \bar{O}^{(s)}_{\beta}$. The compatibility kernel for the cubic/tetragonal transition can then be written as the $ 2 \times 2$ matrix

\begin{equation}
\displaystyle{U_{\ell,\ell'} =\nu (\vec k) \dfrac{G_{\ell \ell'} + \{G_{\ell \ell'} G_{11} - G_{\ell 1}G_{\ell' 1}\} }{1 + G_{11}}}.
\end{equation}
where $\nu (\vec k) \equiv (1 - \delta_{{\vec k}, 0})$ sets the non-OP harmonic-energy contribution for uniform  strains to its minimum value of zero.

\begin{thebibliography}{}
\bibitem{R1} V.K. Wadhawan, {\it Introduction to Ferroic Materials}
(Gordon and Breach, New York, 2000).

\bibitem{R2} K. Bhattacharya, {\sl Microstructure of Martensite} (Oxford University Press, Oxford, 2003); A.G. Khachaturyan, {\sl Theory of structural Transformation in Solids} (Wiley, 1983); J.M.~ Ball and R.D.~James, Phil. Trans. Roy. Soc., Lond. A {\bf 338}, 389 (1992). 


\bibitem{R3} F. Falk, Z. Phys. B {\bf 51}, 177 (1983),  J.-C. Tol\'edano and P. Tol\'edano, {\it The Landau
Theory of Phase Transitions} (World Scientific, Singapore, 1987).

\bibitem{R4} S. H. Curnoe and A. E. Jacobs, Phys. Rev. B {\bf 63}, 094110 (2000).
A. E. Jacobs, S. H. Curnoe and R. C. Desai, Phys. Rev. B {\bf 68}, 224104 (2003); A.E.Jacobs, Phys. Rev, B {\bf 52}, 6327 (1995);
B.Muite and O.U.~Salman, ESOMAT 2009, 03008 (2009) . 



\bibitem{R5}  S. Kartha, T. Castan, J. A. Krumhansl and J. P. Sethna, Phys. Rev. Lett. {\bf 67}, 3630 (1991). 

\bibitem{R6} Y.H. Wen, Y. Wang and L.Q. Chen, Phil. Mag. A {\bf 80}, 1967 (2000).

\bibitem{R7} T. Lookman, S. R. Shenoy, K.O. Rasmussen, A. Saxena and A. R. Bishop, Phys. Rev. B {\bf 68}, 224104 (2003):
K. O.  Rasmussen, T. Lookman, A. Saxena, , A. R. Bishop, R. C.  Albers,  and S. R. Shenoy,
Phys. Rev. Lett. {\bf 87}, 055704,(2001).

\bibitem{R8} C. Manolikas and S. Amelinckx, Phys. Status Solidi A {\bf 61}, 179 (1980); J.W. Seo and D. Schryvers, Acta Materiala {\bf 46}, 1165-1175 (1998).

\bibitem{R9} J. Dec, Phase Trans. {\bf 45}, 35 (1993),  A. L. Roytburd, Phase Trans. {\bf 45}, 1 (1993).

\bibitem{R10} G.R. Barsch, B. Horowitz and J.A. Krumhansl, Phys. Rev. Lett., {\bf 59}, 1251 (1987);  B. Horowitz, G.R. Barsch and J.A. Krumhansl, Phys. Rev. B, {\bf 43}1021(1991).

\bibitem{R11} S.R. Borg, {\it Fundamentals of Engineering Elasticity}, World Scientific, Singapore (1990); E. Kroener in {\it Physics of Defects}, ed R. Balian, M. Kleman and J-P. Pourier, 
Les Houches Session XXV, North Holland (1980).

\bibitem{R12} M. Baus and R. Lovett, Phys. Rev. Lett. {\bf 65}, 1781 (1990); Phys. Rev. A {\bf 44}, 1211 (1991).

\bibitem{R13} R. Ahluwalia, T. Lookman and A. Saxena, Acta Materialia, {\bf 54}, 2109-2120 (2006).

\bibitem{R14} M. Porta, T. Castan, P. Lloveras, T. Lookman, A. Saxena, and S.R. Shenoy, Phys. Rev. B {\bf 79}, 214117, 2009.

\bibitem{R15} P.A. Lindgard and O. Mouritsen, Phys. Rev. Lett., {\bf 57}, 2458 (1980);  A. M. Bratkovsky, S.C. Marais, V. Heine and E.K.H. Salje, J. Phys. Condens. Matt., {\bf 6}, 3769 (1994); E. Vives, J. Goicoechia, J. Ortin and A. Planes, Phys. Rev. E, {\bf 52}, R5 (1995).

\bibitem{R16} B. Cerruti and E. Vives, Phys. Rev. E, {\bf 77}, 064114 (2008); D. Sherrington, J. Phys. CM, {\bf 20}, 304213 (2008).

\bibitem{R17} T. Lookman, S.R. Shenoy and A. Saxena Bull. Am. Phys. Soc., {\bf 49} (1), 1315 (2004);  S.R. Shenoy and T. Lookman, Phys. Rev. B {\bf 78}, 144103 (2008); S.R. Shenoy, T. Lookman and A. Saxena, submitted Phys. Rev. B..

\bibitem{R18} G. R. Barsch and J. A. Krumhansl, Phys. Rev. Lett. {\bf 53}, 1069 (1984); Metall. Trans. A {\bf 19}, 761 (1988).

\bibitem{R19} R. B. Potts, Vol.{\bf 48}, pp. 106-109, (1952); F. Y. Wu, Reviews of Modern Physics, Vo. {\bf 54}, pp. 235–268, (1982). 

%\bibitem{R20} S. R. Shenoy and T. Lookman, Phys. Rev. B {\bf 78}, 144103, (2008).

\bibitem{R20} D. M. Hatch, T. Lookman, A. Saxena, and S. R. Shenoy, Phys. Rev . B {\bf 68},  104105 (2003).


\bibitem{R21}S. Sarkar, X. Ren and K. Otsuka, Phys. Rev. Lett. {\bf 95}, 205702 (2005); R. Vasseur and T. Lookman, Phys. Rev. B {\bf 81}, 094107 (2010).


\bibitem{R22} M. Blume, Phys. Rev. {\bf 141}, 517, (1966); H. W. Capel, Physica (Amsterdam) {\bf 32}, 966, (1966); 
M. Blume, V. J. Emery, and R. B. Griffiths, Phys. Rev. A {\bf 4}, 1071, (1971).

\bibitem{R23} H. Buttner, Y. B. Gaididei, A. Saxena, T. Lookman and A. R. Bishop, J. Phys. A {\bf 37}, 85955-8608, (2004).
A. Saxena and G.R. Barsch, Physica D, {\bf 66}, 195 (1993).

\bibitem{R24}  J. V. Jos\'e, L. P. Kadanoff, S. Kirkpatrick and  D. R. Nelson, Phys. Rev. B, {\bf 16}, 1217 (2007).

\bibitem{R25} J. M. Kosterlitz and D. J. Thouless, J. Phys. C {\bf 6}, 1180 (1973) 
 
\bibitem{R26} M. Rao, S. Sengupta and H.K. Sahu, Phys. Rev. Lett., {\bf 75}, 2164 (1995); K.M. Crosby and R.M. Bradley, Phil. Mag. Lett., {\bf 75}, 131 (1997).

\bibitem{R27} N. Shankaraiah, K.P.N.  Murthy, T. Lookman and S. R. Shenoy, Phys. Rev. B, submitted.


\end{thebibliography}
 \end{document}